\documentclass[12pt,twoside]{article}          
\usepackage{eufrak,amsbsy,latexsym}            

\newcommand{\papernumber}[1]{\hskip 12 true cm #1 \par}

\newcommand{\papertitle}[1]{{\renewcommand{\thefootnote}{\fnsymbol{footnote}}
                       \Large\bf\vskip 0 true cm
                       \begin{center}#1\end{center}
                       \setcounter{footnote}{0}}
                       \normalsize\vskip 2 true cm}
\newcommand{\paperauthor}[1]{{
                       \begin{center} {\large #1 }\end{center}}
                       \setcounter{footnote}{0}}
\newcommand{\paperadress}[1]{\vspace*{-0.8 true cm}\begin{center} {\it #1 }
                         \end{center}
                         \vskip 1.7cm}

\def\CC{{\rm \kern.24em \vrule width.02em height1.5ex depth-.05ex 
        \kern-.26em C}}
\def\HH{{\rm I \kern-.15em H}}
\def\ZZ{{\sf Z \kern-0.46em Z}}

\def\NN{{\rm I \kern-.15em N}}

\def\be{\begin{equation}}
\def\ee{\end{equation}}
\def\ben{\begin{displaymath}}
\def\een{\end{displaymath}}
\def\bea{\begin{eqnarray}}
\def\eea{\end{eqnarray}}
\def\bean{\begin{eqnarray*}}
\def\eean{\end{eqnarray*}}

\def\eqref#1{eq.~(\ref{#1})}

\def\sss{\scriptscriptstyle}
\def\OD{\Omega_{\sss D}}
\def\OM{\Omega_{\sss M}}
\def\cA{{\cal A}}
\def\cAM{\cA_{\sss\cal M}}

\def\cJ{{\cal J}}
\def\cK{{\cal K}}
\def\cM{{\cal M}}
\def\cN{{\cal N}}

\def\dD{d_{\sss D}}
\def\dC{d_{\sss C}}
\def\dMU{d_{\sss \mu}}
\def\dM {d_{\sss\cal M}}
\def\DM{D_{\sss M}}
\def\multD{\odot_{\sss\cal D}}

\def\OCL{\Omega_{\sss D}}

\def\multCL{\odot_{\sss\cal D}}
\def\gFCL{F_{\sss CL}}
\def\gLCL{{L_{\sss CL}}}

\def\OMM{\Omega_{\sss M}}
\def\dMM{d'}
\def\multMM {\bullet}
\def\gFMM{F_{\sss M}}
\def\gLMM{{L_{\sss MM}}}

\def\ON{\Omega_{\sss N}}
\def\dN{d_{\sss N}}
\def\multN{\odot_{\sss N}}
\def\gAN{{A_{\sss N}}}
\def\gFN{F_{\sss N}}

\def\dcN{{d_{\sss \cN}}}

\def\OP{\Omega_{\sss P}}
\def\dP{d_{\sss P}}
\def\multP{\odot_{\sss P}}
\def\multcM{\bullet_{\sss\cal M}}
\def\gFP{{F_{\sss P}}}
\def\gLP{{L_{\sss P}}}
\def\ODM{\Omega_{\sss D_M}}

\def\sotimes{{\raisebox{.35ex}{$\sss\otimes$}}}
\def\soplus{{\raisebox{.35ex}{$\sss\oplus$}}}

\def\gA{{A}}
\def\gF{{F}}

\newcommand{\xker}{\mbox{ker}}

\def\kin{\,\in\,}

\begin{document}

\papernumber{MZ-TH/95-11}

\papertitle{ Natural extensions of the Connes--Lott Model and
   comparison with the Marseille--Mainz Model\footnote{
   Work supported in part by PROCOPE project Mainz University and CPT
   Marseille--Luminy.}
 }

\paperauthor{ N.A. Papadopoulos\footnote{ E--mail:
   papadopoulos@dipmza.physik.uni--mainz.de}, J. Plass } \paperadress{ Institut
   f\"{u}r Physik,\\ Johannes Gutenberg Universit\"{a}t\\
   Staudingerweg 7, D--55099 Mainz, Germany }

\begin{abstract}
An extension of the Connes--Lott model is proposed. It is also
within the framework of the A.Connes construction based on a generalized
Dirac--Yukawa operator and the K--cycle $(H,D)$, with $H$ a fermionic
Hilbert space.
The basic algebra $\cA$ which may be considered as representing the 
non--commutative extension, plays a less important role in our approach. 
This allows a new class of natural extensions. The proposed extension lies 
in a sense between the Connes--Lott and the Marseille--Mainz model.
It leads exactly to the standard model of electroweak interactions.
\end{abstract}

\newpage

{\bf 1.}\hspace{1em}
In this paper we present an extension of the model developed by
A.~Connes and J.~Lott [1--3]. 
In some sense the proposed extension lies between the Connes--Lott model
and the Marseille--Mainz model
[4--7] 
within the noncommutative geometry approach to the standard model in
elementary particle physics.

One of the motivations for the proposed generalisation is to avoid
some peculiar features of the Connes--Lott model (CL model).
These include the presence of a $\gamma^5$ matrix in Yukawa
couplings, unusual in the standard model, the assignment of the fermionic
additive quantum numbers and the absence of spontaneous symmetry
breaking for one generation of fermions.
In addition, we propose an alternative construction whose spirit is close to
the one of Connes' and Lott's construction but avoids some of their
predictions \cite{3} which may be in contradiction with experiment.
Finally we stress the role of
the basic associative algebra $\cA$ in the CL model and its phenomenological
implications.

The following observation may be considered as the physical motivation for the
approach based on noncommutative geometry. As for the electromagnetic 
interactions the photon related to the Abelian group $U(1)$ leads to the 
Minkowski
space--time, the electroweak interactions, and especially the W and Z particles
which are connected with a noncommutative group $SU(2)\times U(1)$, may lead to
a space--time where noncommutative geometry plays an important role. One of the
main virtues of the noncommutative geometry approach is the explanation for the
spontaneous symmetry breaking and the Higgs effect in the standard model.

A common aspect of the above models, inspired by noncommutative geometry, is
the construction and importance of a certain graded differential (or derivative)
algebra $\Omega^*$ which may be considered as "noncommutative" generalisation
of the de Rham algebra $\Lambda^*(X)$ of differential forms over
space--time X, and which is the space where the gauge potential $\gA$ lives. 
The models differ from each other in the choice of the specific graded 
differential
(or derivative) algebra $\Omega^*$ they use for the construction of the gauge
potential (super--connection) $\gA$ and field strength (super--curvature)
$\gF$.
This leads to somewhat different versions of the standard model in the two
cases.
There is indeed such a difference between the Marseille--Mainz model (MM model)
and the CL model: The MM model leads exactly to the standard model with
spontaneous symmetry breaking and to the Higgs potential, and gives a natural
framework for the discussion of the CKM matrix
\cite{8}
but it does not determine any of its parameters. The reason for this is
explained and discussed in
\cite{6}.
The CL model, which contains more structure, seems to lead not exactly
to the standard model but to a variation of it. This
follows from the fact that, with one generation only, this model shows no
spontaneous symmetry breaking.
It is not clear whether the presence of $\gamma^5$ in the
Higgs sector leads to some unusual coupling. The quantum
numbers of fermions are not from the beginning equivalent to those of the
standard model unless an additional Poincare duality assumption enters the
construction
\cite{2,3,10}.
The results of \cite{6,7} may suggest that it is at least questionnable
whether the CL model can fix some parameters of the standard
model \cite{3,10}, even at the classical level.
In what follows, we first give
a brief review of the CL and the MM model (section 2 and section 3),
and then proceed with the proposed extension of the CL model
(sections 4, 5 and 6).

{\bf 2.}\hspace{1em}
Since we would like to extend the CL model, it is necessary to first review
some of its aspects.
Here this is formulated in much simpler terms than in the original version
\cite{1,2,12}.
This is possible by exploring the mathematical results of
\cite{11}.
This was also demonstrated in
\cite{7}
with a toy model. Here we use for the first time the new formulation in the
realistic case. In addition, we discuss the role of the basic
algebra $\cA$ which constitutes the most
essential difference between the CL and the MM models, and which is also the
starting point for the proposed generalisation.

We essentially follow the lines of
\cite{7}
and 
\cite{11}
but choose the associative algebra  to be
\ben
  \cA \,=\, \cAM \otimes C^\infty(X)
\een
with
\ben
  \cAM \,=\, \HH \oplus \CC 
\een
instead of $\cAM = \CC \oplus \CC$ in
\cite{7}
.
$\HH$ represents the quaternionic numbers and $C^\infty(X)$ the smooth 
functions on the space--time $X$. This algebra leads to a standard model
like version.
 
Given the associative algebra $\cA$, one considers
first its universal differential envelope
   $(\Omega^*(\cA),\delta)$,
which is generated by the formal elements ("words")
   $A_0\delta A_1 \ldots \delta A_n \kin \Omega^n(\cA)$
and the operator 
   $\delta$
obeying the Leibniz rule
   $\delta(AB) = (\delta A)B + A(\delta B)$.
By means of a K--cycle (Dirac--Kasparov cycle) $(H,D)$ over
$\cA$, consisting of a Hilbert space $H$,
a Dirac--Yukawa operator $D$ on $H$,
and a representation of $\cA$ on $H$, the associative algebra
$\Omega^*(\cA)$ is represented on the space $L(H)$ of bounded
linear operators over $H$ by
\ben
   \pi\,:\, \Omega^*(\cA) \longrightarrow L(H) \,,\,
   A_0\delta A_1 \cdots \delta A_n \longrightarrow A_0[D,A_1]\cdots [D,A_n].
\een
The Dirac--Yukawa operator has the form
\be\label{eq-7}
   D=i\gamma^\mu\partial_\mu +\DM \quad,\quad \DM=\eta
\ee
with $\eta$ a matrix as specified below (see \eqref{eq-3}).
$\DM$ may be understood to be a fermionic mass matrix. Note that 
such an interpretation was not made in the MM approach because it
is unnecessary in that framework.
In the original version of Connes' and Lott's construction, the 
gauge potential and the field strength were taken to be elements 
of $\pi(\Omega^*(\cA))$.
Since the representation $\pi$ fails to respect the differential 
structure of $\Omega^*(\cA)$,
this leads to the appearance of auxiliary or adynamic fields  
(fields without kinetic energy) in the Lagrangian which have to 
be eliminated by minimization 
\cite{1,2,13,14}.
At this stage a direct comparison with other approaches, such as
the MM model, is not possible.
In the more recent version of the CL model given in
\cite{12}
and later in
\cite{3},
one considers in addition the space $\OD^*(\cA)$,
obtained from $\Omega^*(\cA)$  by dividing out the ideals
$\cJ^k(\cA)=\cK+\delta\cK^{k-1}$, where
$\cK^k := \xker\pi \cap\Omega^k$:
\ben
   \OD^k(\cA) \,=\, \Omega^k(\cA)      \,/\, \cJ^k(\cA)
   \quad,
\een
or equivalently
\be\label{eq-6}
   \OD^k(\cA) \,=\, \pi(\Omega^k(\cA)) \,/\, \pi(\cJ^k(\cA))
   \quad.
\ee

In contrast to $\pi(\Omega^*(\cA))$, the space  $\OD^*(\cA)$
is an $\NN$--graded differential algebra (like the universal object
$\Omega^*(\cA)$). Therefore $\OD^*(\cA)$ is the space which 
should be compared to the space $\OM^*(X)$ in the MM model
(see below). The multiplication law is defined by the ordinary
multiplication in $L(H)$ and by taking the quotient. We denote
it by the symbol $\multD$. Similarly the differential $\dD$ on
$\OD^*(\cA)$ is defined by means of commutators with the 
Dirac--Yukawa operator and by taking the quotient as above. 
The structure of this algebra may therefore be summarized as follows:
\ben
   (\,\OD^*(\cA)\,,\,\multD\,,\,\dD\,)
\een
The explicit construction of the space $\OD^*(\cA)$ was given in
\cite{11}.
Since this result is particularly important for the treatment
below, we would like to give a short discussion of it
\footnote{
   It is indeed this result which allows our simplified treatment
   and the direct comparison between the two models (CL and MM),
   and it also gives a hint for the extension of the CL model
}.
In the case where basic algebra is of the form
$\cA = \cAM \otimes C^\infty(X)$
with $\cAM$ a block diagonal matrix algebra, the differential algebra
\be\label{eq-4}
   (\,\OD^*(\cA)\,,\,\multD\,,\,\dD\,)
\ee
is isomorphic to the skew tensor product of the de Rham algebra
$(\Lambda^*(X),\dC)$ and a specific (quotient space) matrix
differential algebra
$\mu^*$  which is $\NN$--graded and generated from $\cAM$\cite{11}:
\be\label{eq-5}
   \OD^*(\cA) \,=\, \mu^*(\cA) \,\hat\otimes\, \Lambda^*(X)
   \quad,\quad
   \OD^k(\cA) \,=\, \bigoplus^k_{i=0} \, \mu^{k-i}\otimes\Lambda^i(X)
\ee
The total grade of homogeneous elements
$[a]\sotimes\,\alpha$ with $[a]\in\mu^*$ and $\alpha\in\Lambda^*$
is given by
\ben
   \partial([a]\otimes\alpha)
   \,=\,
   \partial([a]) \,+\, \partial(\alpha) 
   \quad .
\een
The multiplication law in $\OD^*(\cA)$  reads
\be\label{eq-1}
   ([a]\otimes\alpha)\multD([b]\otimes\beta)
   \,=\,
   (-1)^{\partial([b])\partial(\alpha)}
   [a][b]\otimes\alpha\beta
   \quad .
\ee
The differential $\dD$ is given by
\be\label{eq-2}
   \dD([a]\otimes\alpha)
   \,=\,
   \dMU[a]\otimes\alpha \,+\, (-1)^{\partial[a]}[a]\otimes\dC\alpha
   \quad .
\ee
It is important to realize that
$(\OD^*(\cA),\multD,\dD)$
depends in an essentiall way on the basic algebra $\cA$ (and of course on $H$
and $D$)
and is uniquely determined once $\cA$ is given.
For the case $\cAM=\CC\oplus\CC$, the calculation was given in
\cite{7}
. Here we present the explicit result for the case $\cAM=\HH\oplus\CC$ which
corresponds to the CL model
[1--3]
. We consider only one generation of fermions.
For the purpose of physics, we need to know only the spaces $\pi(\Omega^k)$
for $k=0,1,2$. Thus we have to determine the projected ideals $\pi(\cJ^k)$
for these three values of $k$. They are found to be
(see also \cite{15})
\ben
   \pi(\cJ^0) \,=\, \{0\}
   \,,\quad
   \pi(\cJ^1) \,=\, \{0\}
   \,,\quad
   \pi(\cJ^2) \,=\, (\CC_{\sss 2\times 2}
                     \oplus
                     \CC)\otimes\Lambda^0(X)
   \quad.
\een
Using
\be\label{eq-12}
   M^0 \,:=\, \cAM \,=\,
   \left(
      \begin{array}{cc}
         \HH & 0    \\
         0   & \CC
      \end{array}
   \right)
   \,,\,\,
   M^1 \,:=\,
   \left(
      \begin{array}{ccc}
         0   & 0   & \CC    \\
         0   & 0   & \CC    \\
         \CC & \CC & 0
      \end{array}
   \right)
   \,,\,\,
   M^2 \,:=\,
   \left(
      \begin{array}{cc}
         \CC_{\sss 2\times 2}   & 0      \\
         0                      & \CC
      \end{array}
   \right)
   \,,
\ee
we obtain, in an obvious notation,
\bean
   \OD^0 &=& \cA \,=\, M^0\sotimes \Lambda^0
             \quad, \\
   \OD^1 &=& (M^0\sotimes \Lambda^1)
             \,\soplus\,
             (M^1\sotimes \Lambda^0) 
             \quad, \\
   \OD^2 &=& (M^0\sotimes \Lambda^2) 
             \,\soplus\,
             (M^1\sotimes \Lambda^1)
             \,\soplus\,
             (M^2\sotimes \Lambda^0)/(M^2\sotimes \Lambda^0)
     \,\, = \,\, (M^0\sotimes \Lambda^2) 
             \,\soplus\,
             (M^1\sotimes \Lambda^1)
             \quad.
\eean
The multiplication law can easily be derived 
for the case $\OD^1\times\OD^1\rightarrow\OD^2$:
\ben
   \left( (M^0\sotimes \Lambda^1)\soplus(M^1\sotimes \Lambda^0) \right)
   \,\multCL\,
   \left( (M^0\sotimes \Lambda^1)\soplus(M^1\sotimes \Lambda^0) \right)
   \,=\,
   (M^0\sotimes \Lambda^2)\soplus(M^1\sotimes \Lambda^1)
   \quad .
   \nonumber
\een
In particular it is important to note that
\ben
   (M^1\sotimes \Lambda^0)
   \,\multCL\,
   (M^1\sotimes \Lambda^0)
   \,=\,
   0
   \quad .
   \nonumber
\een
Similarly we obtain for the differential $\dD:\, \OD^1\rightarrow\OD^2$:
\ben
   \dD\left((M^0\sotimes \Lambda^1)\soplus(M^1\sotimes \Lambda^0)\right)
   \,=\,
   (M^0\sotimes\dC\Lambda^1)
   \soplus
   (-M^1\sotimes\dC\Lambda^0+\dM M^0\sotimes\Lambda^1) \\
   \soplus
   (\dM M^1\sotimes \Lambda^0)
   \quad .
   \nonumber
\een
with
\ben
   \dM \left(M^0\sotimes\Lambda^1 \right) \,=\, [\eta,M^0]\sotimes\Lambda^1
   \quad \textrm{and}\quad
   \dM \left(M^1\sotimes\Lambda^0 \right) \,=\, \{\eta,M^1\}\sotimes\Lambda^0
   \quad ,
   \nonumber
\een
$[,]$ representing the commutator, $\{,\}$ representing the anticommutator
and
\be\label{eq-3}
   \eta \,=\,
   i\left(
      \begin{array}{cc}
         0              & C \\
         \overline{C}   & 0
      \end{array}
   \right)
   \quad,\quad 
   C = {1 \choose 0}
   \quad .
\ee
Note that because of the multiplication law $\multD$ we obtain
$\{\eta,M^1\sotimes\Lambda^0\}_{\sss \multD}=0$.

The generalized potential (super--connection) $\gA$ is a (skew hermitian)
element of $\OD^1(\cA)$ and reads explicitly:
\bea\label{eq-8}
   \gA \,=\,
   i \left(
      \begin{array}{cc}
         A_{\sss SU(2)}   & \Phi \\
         \overline{\Phi} & B_{\sss U(1)} 
      \end{array}
   \right)
   && \textrm{with } A_{\sss SU(2)} =\frac{1}{2} \tau_i A^i_\mu dx^\mu
            \,,\quad B_{\sss U(1)} = B_\mu dx^\mu\quad,  \\
   && \textrm{and } \Phi \textrm{ the scalar field } 
                   \Phi= {\Phi^0\choose \Phi^-} \nonumber
   \quad .
\eea
The structure group is $SU(2)\times U(1)$ given by the unitary part of $\cAM$.
The field strength (super--curvature) is obtained from the structure equation
\ben
    \gFCL \,=\, \dD\gA \,+\, \gA\,\multD\, \gA
    \quad.
\een
A straightforward calculation, using the multiplication rule and the
differential
given above, leads to the result
\bea
   \gFCL \,=\,
   i \left(
      \begin{array}{cc}
         F^{\sss A}     & -D\Phi \\
         -\overline{D\Phi}            & F^{\sss B} 
      \end{array}
   \right) \nonumber
   && \textrm{with } F^{\sss A} =
                       \dC A_{\sss SU(2)}+A_{\sss SU(2)}\wedge A_{\sss SU(2)} 
                     \,,\,\,
                     F^{\sss B} = \dC B_{\sss U(1)} \\
   && \textrm{and } D\Phi = \dC\Phi + A_{\sss SU(2)}(\Phi+C)
                             -B_{\sss U(1)}(\Phi-C) \label{eq-10}
                    \,\, .
\eea
The Lagrangian $\gLCL=-tr\gFCL^+\gFCL$ is given by
\ben
\gLCL \,=\, -\frac{1}{4}\, tr F^{\sss A}_{\mu\nu}  
                      F^{{\sss A}\mu\nu}
         \, -\frac{1}{4}\,    F^{\sss B}_{\mu\nu}  
                      F^{{\sss B}\mu\nu}
         \, + 2\overline{D\Phi}D\Phi
         \quad.
\een
From this it is obvious that the Higgs potential in the CL model with one
generation
of fermions (leptons) is trivial 
$V^{\sss \Phi}_{\sss CL} =0$\,.

{\bf 3.}\hspace{1em}
For the benefit of the reader but also in order to facilitate the comparison of
the various models, we would like to give a very short review of some
essential ingredients of the MM model [4-7] 
before starting with the extension of the CL model. 

The starting point of the MM model is the $\ZZ_2$--graded algebra
$\OM^*(X)$ obtained from the skew tensor product of the matrix algebra
$\CC_{\sss 3\times 3}$ and the algebra $\Lambda^*(X)$ of
differential forms over the space--time $X$. The matrix algebra
$\OM^*(X)=\CC_{\sss 3\times 3}\hat\otimes\Lambda^*(X)$
is taken $\ZZ_2$--graded with $\Gamma=diag(1,1,-1)$ as the grading
automorphism.
\newline
The matrix multiplication and generalized differential $\dMM$ is,
with an obvious change of notation and with $a$ instead of $[a]$,
given formally as in
\eqref{eq-1} and \eqref{eq-2} respectively.
 The matrix derivative $\dM$ in $\CC_{\sss 3\times 3}$
is defined by its action on the even and odd part of $a_0$ and $a_1$
respectively
\cite{6}, with $\eta$ given as in
\eqref{eq-3}
:
\ben
   \dM(a) \,=\, [\eta,a_0] \,+\, i\{\eta,a_1\}
   \quad .
\een
The structure of the algebra may be summarized by
\ben
   (\, \OMM^*\,,\, \multMM \,,\, \dMM \,)
\een
It should be understood that in the MM model no quotient space is present so
that the $(\OMM^*)$ multiplication $\multMM$ and differential $\dMM$
are induced straightforwardly from the tensor structure and of course are
much simpler than in the CL model.
It is important to realize that $(\OMM^*,\multMM,\dMM)$ is not a differential
algebra since $\dMM$ is only a derivation and not a differential. It is
interesting, however, to note that $\OMM^1=\OD^1$ is valid. So we may start 
with the same gauge potential $\gA$ as in
\eqref{eq-8}
in both cases. For the field strength we have in the MM model
\ben
    \gFMM \,=\, \dMM\gA \,+\, \gA\,\multMM\, \gA
    \quad.
\een
The Lagrangian for the bosonic part is given by
\cite{4,5}
\ben
    \gLMM \,=\, \gLCL \,-\, V^{\sss \Phi}
    \quad
    \textrm{with }
    V^{\sss \Phi} 
      = 2{(\overline{\Phi}C+\overline{C}\Phi+\overline{\Phi}\Phi)}^2
    \quad.
\een
In the MM model, we obtain a non trivial Higgs potential even with one
generation of fermions in accordance with the standard model.
This is an important difference with the CL model.

{\bf 4.}\hspace{1em}
With the above preparation we are in the position to formulate our new model in
a precise and, as we hope, efficient way. The most important difference
between the CL and the MM model is the importance of the basic algebra
$\cA$. In the CL model, the entire construction relies on the algebra
$\cA$. In addition, the algebra $\cA$ has also direct phenomenological
implications. Not only the determination of the relevant structure group
$SU(2)\times U(1)$ but also the determination of the relevant fermion
representations and in particular the fermion quantum numbers are derived from
the associative algebra $\cA$. In the CL model one uses the particular
differential algebra $\OD^*(\cA)$ since one started with
$\cA=\cAM\otimes C^\infty(X)$.
So it is this particular $\cA$ which fixes the gauge potential $\gA$ as an 
element in $\OD^1(\cA)$.
In addition it is this $\cA$ which determines completely the
fermionic part
of the Lagrangian including all phenomenological implications
\cite{3}.

In the MM model, the starting point is a derivative algebra
$\Omega^*$ which generalizes
the algebra $\Lambda^*(X)$
\footnote{
   This aspect is common also to other models within the noncommutative
   geometry approach
   \cite{9}.
}.
The gauge potential $\gA$ is an element of $\Omega^1$. We started with
$\CC_{\sss 3\times 3}$ for the construction of $\OMM^*$
only because the representation space of the Lie algebra of 
$SU(2|1)$ is a $\CC^3$. Here it is the super Lie algebra $SU(2|1)$ only
which leads to the phenomenological consequences. The result is that the MM
model, as is well known
\cite{6},
even if it has the nice feature, among others, to explain e.g. spontaneous
symmetry breaking and the Higgs effect in a geometrical way, has less
predictive power in the fermionic sector and gives exactly the standard model.
The CL model, because of the fundamental role played by the associative
algebra $\cA$, seems to have more predictive power
\cite{3}.

At this point, some comments are appropriate. The fundamental role played by
an associative algebra is a new aspect of the phenomenology in elementary
particle physics. Usually Lie algebras are used for phenomenological
implications because they correspond to infinitesimal symmetry transformations.
Associative algebras are used mathematically to define the corresponding Lie
algebra by the use of commutators. Furthermore, it is well known that in
current algebra, the commutator product which determines the Lie
algebra is well defined whereas the associative product itself is not well
defined (it may be infinite). Therefore it should be stated that the use of an
associative algebra for the phenomenology is not at all a priori well justified
from the physical point of view. It is an essential ingredient and an
important and basic aspect of the CL model which distinguishes this model
from all other models within the framework of noncommutative geometry in
elementary particle physics.

{\bf 5.}\hspace{1em}
If we relax this fundamental role of the associative algebra $\cA$,
we obtain a new class of models $\{N\}$ which rely essentially only on the
differential algebra $\{\ON^*\}$ which contains also the algebra $\OD^*$.
We use $\cA$ only as an instrument for the mathematical
construction of the new differential algebra $\ON^*$ and for nothing more. We
use $\ON^1$ to determine the gauge potential $\gAN$ and $(\ON^*,\multN,\dN)$
to derive the field strength $\gFN$:
\ben
   \gFN \,=\, \dN\gAN\,+\, \gAN\multN\gAN
   \quad.
\een
The bosonic part of the Lagrangian is obtained by taking
$tr(\gFN^+\gFN)$. No specific predictions are made for the fermionic part.

A natural extension of the CL model is obtained by taking a new
differential algebra
\ben
    (\,\ON^*\,,\,\multN\,,\,\dN\,)
\een
which generalizes the expression
\eqref{eq-5}
for $\OD^*$:
\ben
   \ON^*=\cN^*\,\hat\otimes\,\Lambda^*(X)
\een
$\ON^*$ is a skew tensor product of a differential matrix algebra
$(\cN^*,\dcN)$ and the differential forms $\Lambda^*(X)$. It is formally fixed
by the analogous expression which follows
\eqref{eq-5}.
Index $D$ is replaced by index $N$ and $\mu^*(\cA)$ by $\cN^*$.
Since no specific restriction is made for $\cN^*$
the algebra $\OD^*$ in the CL model is a special case of the
differential algebra $\ON^*$.

We choose now another special case of $\ON^*$ which is very near to the spirit
of the CL model. This will give the concrete new model we would like to
discuss. We denote this special differential algebra by $\OP^*$ and we have in
an obvious notation, following
\eqref{eq-4}, \eqref{eq-5}:
\ben
   (\,\OP^*\,,\,\multP\,,\,\dP\,)
   \quad\textrm{ with }\,\,
   \OP^*=\cM^*\,\hat\otimes\,\Lambda^*(X)
   \quad.
\een

The matrix algebra $\cM^*$ depends on $\cAM$, \eqref{eq-12},
and $\DM$, \eqref{eq-7}, and is given by
\be\label{eq-9}
   \cM^k \,=\, \ODM^k(\cAM) \,=\, \pi(\Omega^k(\cAM))/\pi(\cJ^k(\cAM))
   \quad.
\ee
We used the notation of
\eqref{eq-6}.
It is obvious that in our model the division in $\pi(\Omega^*)$ does not
depend on the $C^\infty(X)$ part of $\cA$. The division concerns only the
matrix space $\cAM$ we started with. We also choose here $\cAM=\HH\oplus\CC$
in order to obtain the right structure group $SU(2)\times U(1)$. This is the
only reason for that choice and we do not fix anything else in
the fermionic sector.

Our next step is to determine the space $\OP^*$. For that purpose we proceed in
a slightly different way than in the CL model (for the $\OD^*)$ and we first
determine the space $\cM^*$. The construction of $\OP^*$ is
obtained in a straightforward manner by the skew tensor product of $\cM^*$ and
$\Lambda^*$.

Using $\DM$ as in \eqref{eq-7}, and $C$ as in \eqref{eq-3}, we obtain
\ben
   \pi(\Omega^{2k}(\cAM)) =
   \left(
      \begin{array}{cc}
         \CC_{\sss 2\times 2} & 0 \\
         0                    & \CC
      \end{array}
   \right)\textrm{ for } k\ge 1 \,,\,\,
   \pi(\Omega^{2k+1}(\cAM)) =
   \left(
      \begin{array}{cc}
         0               & \CC C \\
         \CC\overline{C} & 0
      \end{array}
   \right)\textrm{ for } k\ge 0
\een
and
\ben
   \pi(\cJ^{0}(\cAM)) \,=\, \{0\}
   \,,\quad
   \pi(\cJ^{1}(\cAM)) \,=\, \{0\}
   \,,\quad
   \pi(\cJ^{2}(\cAM)) \,=\,
   \left(
      \begin{array}{cc}
         i\HH & 0 \\
         0    & 0
      \end{array}
   \right)
   \,,
\een
\ben
   \pi(\cJ^{k}(\cAM)) \,=\, \pi(\Omega^{k}(\cAM)) \quad\textrm{for } k\ge 3
   \quad.
\een
So we obtain from
\eqref{eq-9}
\ben
   \cM^0 =
   \left(
      \begin{array}{cc}
         \HH & 0   \\
         0   & \CC
      \end{array}
   \right)
   \,,\,
   \cM^1 =
   \left(
      \begin{array}{cc}
         0               & \CC C \\
         \CC\overline{C} & 0
      \end{array}
   \right)
   \,,\,
   \cM^2 =
   \left(
      \begin{array}{cc}
         \HH & 0   \\
         0   & \CC
      \end{array}
   \right)
   \,,\,
   \cM^k = \{0\} \textrm{ for } k\ge 3
   \,.
\een
The multiplication $\multcM$ and the differential $\dM$ are given canonically
by the quotient of $\pi(\Omega^*(\cAM))$ by $\pi(\cJ^*(\cAM))$ and we have in 
an obvious notation:
\ben
   [a]\multcM [b] \,:=\, [ab] \quad\textrm{ and }\quad 
   \dM[a] \,=\, [\DM,a]
   \quad.
\een
The space $(\OP^*,\multP,\dP)$ is given by
\ben
   \OP^j \,=\, (\cM^0 \sotimes \Lambda^j)
               \soplus
               (\cM^1 \sotimes \Lambda^{j-1})
               \soplus
               (\cM^2 \sotimes \Lambda^{j-2})
   \quad.
\een
So we have explicitly for $j=0,1,2$:
\be\label{eq-11}
   \OP^0 = \cA
             \,,\,\,
   \OP^1 = (\cM^0 \sotimes \Lambda^1)
             \soplus
             (\cM^1 \sotimes \Lambda^0)
             \,,\,\,
   \OP^2 = (\cM^0 \sotimes \Lambda^2)
             \soplus
             (\cM^1 \sotimes \Lambda^1)
             \soplus
             (\cM^2 \sotimes \Lambda^0)
   \,.
\ee

It is interesting to note that
$ \OP^0\,=\,\OCL^0$
and
$\OP^1\,=\,\OCL^1$.
This allows to start with the same super--potential $\gA$ 
as in the CL and MM model (see \eqref{eq-8}) .
The multiplication rule $\multP$ and the differential $\dP$ are now different.
The field strength in this model reads:
\ben
   \gFP \,=\, \dP\gA \,+\, \gA\multP\gA
   \quad.
\een
Using the results of
\eqref{eq-11}
we easily obtain in the notation of 
\eqref{eq-10}
\bean
   \gFP \,=\,
   i \left(
      \begin{array}{cc}
         F^{\sss A}
         -(\Phi\overline{C}+C\overline{\Phi}+\Phi\overline{\Phi})
         & -D\Phi \\
         -\overline{D\Phi} & F^{\sss B} 
                             -(\overline{\Phi}C+\overline{C}\Phi
                             +\overline{\Phi}\Phi)
      \end{array}
   \right)
   \quad.
\eean
This leads to the Lagrangian
\ben
    \gLP\,=\, \gLCL \,-\, V^{\sss \Phi}
    \quad\textrm{with}\quad
    V^{\sss \Phi} 
      = 2{(\overline{\Phi}C+\overline{C}\Phi+\overline{\Phi}\Phi)}^2
    \quad .
\een
It is obvious that this Lagrangian, in contrast to the $\gLCL$, leads to
spontaneous symmetry breaking even with one generation of fermions.
It is also important to realize that at this level,
$\gLP$ is similar to the one in the MM model, although there the
structure group was the group $SU(2|1)$.

{\bf 6.}\hspace{1em}
We may now summarize our results. We have constructed a new model
within the spirit of the Connes--Lott approach which gives precisely
the standard model in elementary particle physics.
It is directly formulated in Minkowski space--time. It contains
none of the unusual aspects of the Connes--Lott model as e.g. the
fixing of the gauge group and consequently the problem of the fixing
of fermionic quantum numbers, the non--existence of spontaneous
symmetry breaking in the case of one generation of fermions and
perhaps the presence of some unusual couplings.

Although formulated in the framework of a differential algebra ($\dP\dP=0$),
it is interesting to note that the bosonic sector of the proposed model is
equivalent to the Marseille--Mainz model, which
is based on an algebra with derivative only ($\dMM\dMM\neq 0$).

We thank R.~H\"au{\ss}ling, W.~Kalau, T.~Sch\"ucker, M.~Walze, J.M.~Warzecha
for discussions and F.~Scheck for reading the manuscript and discussions.

\end{document}